\documentclass[12pt]{article}
\usepackage{epsfig}
\usepackage{amstex}
\usepackage{amssymb}
\usepackage{amsfonts}
\usepackage{array}
\usepackage{theorem}
\usepackage{amstext}

\newtheorem{Lem}{Lemma}
\newtheorem{Cor}{Corollary}
\theoremstyle{break}
\theoremheaderfont{\normalfont\bfseries}
\def\super       {\hat{sl}(2|1;\mathbb C)_k}
\def\iso         {\hat{sl}(2;\mathbb C)}
\def\isop        {\hat{sl}(2;\mathbb C)_{-\frac{1}{2}}}

\def\defd        {\stackrel{\rm d}{=}}
\def\half        {\textstyle{1\over2}}
\def\quarter     {\textstyle{1\over4}}
\def\zplus       {\mathbb Z _{+}}
\def\N           {\mathbb N}

\def\scon        {superconformal }
\def\ns          {Neveu--Schwarz }
\def\r           {Ramond }
\def\Z           {\mathbb Z}
\def\R           {\mathbb R}
\def\C           {\mathbb C}
\def\ben         {\begin{equation}}
\def\een         {\end{equation}}
\def\bea         {\begin{eqnarray}}
\def\eea         {\end{eqnarray}}
\def\nn          {\nonumber \\ }
\newcommand{\im}{\mathop{\mathrm{Im}}}
\def\hslc        {\hat{sl}(2|1;{\mathbb C})}
\def\hslck       {\hat{sl}(2|1;{\mathbb C})_k}
\def\hslcp       {\hat{sl}(2|1;{\mathbb C})_{-\frac{1}{2}}}
\def\hf          {\frac{1}{2}}

\begin{document}
\def\theequation{\thesection.\arabic{equation}}
\begin{titlepage}
\begin{flushright}
DTP/97/19\\ 
May 1997\\
\end{flushright}
\vspace{1cm}
\begin{center}
{\Large\bf Characters of admissible representations of the affine superalgebra $\super$ }\\
\vspace{1cm}
{\large
P.~Bowcock, M.~Hayes and A.~Taormina}\\
\vspace{0.5cm}
{\it Department of Mathematical Sciences, University of Durham,\\
Durham DH1~3LE, England} \\
\vspace{2cm}
\end{center}

%
%
%
%
\begin{abstract}
We calculate characters and supercharacters for irreducible, admissible representations of the
affine superalgebra $\super$ 
in both the Ramond and Neveu--Schwarz sectors and discuss their modular properties in the special case of level $k=-\hf$. We also show that the non-degenerate integrable $\super$ characters coincide with some $N=4$
superconformal characters. 
\end{abstract}
\vskip 8truecm
{}e-mail: Peter.Bowcock@@durham.ac.uk, M.R.Hayes@@durham.ac.uk, Anne.Taormina@@durham.ac.uk
\end{titlepage}
\section{Introduction.}
\renewcommand{\theequation}{1.\arabic{equation}}
\setcounter{equation}{0}

The classical simple complex Lie superalgebra $A(1,0)$ has a realisation
provided by $sl(2|1;\C)$, the set of $3 \times 3$ complex matrices  
whose supertrace is zero. The affinisation of one of its real forms, $sl(2|1;\R)$, plays a central r$\hat{{\rm o}}$le
in the construction of the topological gauged $SL(2|1;\R)/SL(2|1;\R)$ Wess-Zumino-Novikov-Witten (WZNW)
model, which is believed to be intimately related to the non-critical
$N=2$ string theory. In order to substantiate the evidence,
based on similar observations made in \cite{HY,AHA,FY} for the non-critical bosonic and $N=1$ strings, one should calculate the space of physical states
in both theories. Our paper is a contribution towards this task. 
Indeed, the partition function of the $SL(2|1;\R)/SL(2|1;\R)$ theory splits
in three sectors : a level $k$ and a level $-(k+2)$ WZNW models based on
$SL(2|1;\R)$, as well as a system of four fermionic and four bosonic ghosts
corresponding to the four even and four odd generators of $SL(2|1;\R)$
(see e.g.\cite{KAR,AHA}). The physical states of the theory should be obtained as elements of the cohomology of the BRST charge. A crucial step in
this approach \cite{BOUW} is to pass from the cohomology on the Fock space
to the irreducible representations of $\hslc$ at level $k$.
When the matter coupled to supergravity in the $N=2$ non-critical string
is minimal, i.e. taken in an $N=2$ super Coulomb gas representation with
central charge
\ben
c_{\text{matter}}=3(1-\frac{2p}{u}),~~~~,p, u \in \N,~~~{\rm gcd}(p,u)=1,
\end{equation}
the level of the `matter' affine superalgebra $\hslc$ appearing in the 
\newline $SL(2|1;\R)/SL(2|1;\R)$ model is of the form
\ben
k = \frac{p}{u} - 1,
\label{levelp}
\end{equation}
i.e. the level precisely takes the values for which admissible representations of $\hslck$ 
exist \cite{KW88}. It is therefore interesting to study the representation
theory of this affine superalgebra in the context of the $N=2$ non-critical 
string, but also in connection with a large class of applications,
from conformal field theory to integrable systems.

The Kac-Kazhdan determinant formula for $\hslck$ \cite{DOB2,K86} encodes crucial information
on the singular vectors of the theory. In \cite{BT}, 
four classes of representations
and the corresponding embedding diagrams were given, each of them having infinitely
many singular vectors in the highest weight Verma module. 
In this paper, we present character formulas for a subset of the above classes,
which correspond to integrable and admissible representations.
It turns out that for integer values of the level $k$, the character formulae 
coincide with those of the superconformal algebra $N=4$, calculated in \cite{ET881,ET882}. We offer some comments on this rather surprising fact 
later in the paper.

When the level is fractional and of the form \eqref{levelp}, admissible representations do occur for the embedding diagrams of class IV (in the notations of \cite{BT}), which are identical in structure to the embedding diagrams for the minimal representations of the $N=2$ \scon algebra as found in
\cite{DORZ} or in \cite{EGAB}. This analogy between embedding diagrams is reminiscent of the analogy between $\hat{sl}(2,\C)$ (resp.
$\hat{osp}(1|2;\C)$) and Virasoro (resp. $N=1$)
diagrams, and is correlated to the relation between
$\hslck$ and $N=2$ theories through hamiltonian reduction \cite{BO,SEMI}.

The paper is organised as follows. The character formulae corresponding
to class IV embedding diagrams are given in section 2 for arbitrary level
$k$. It is also explained there
that these characters only form a finite representation of the modular group
when class V characters, corresponding to representations whose highest weight
state quantum numbers take values at the edge of the class IV domain, are included. 
In section 3, the characters of integrable representations stemming from class I embedding diagrams are presented and compared with massive $N=4$ superconformal
characters, while the unique class IV integrable character coincides
with the massless vacuum $N=4$ character. Analogies between
the two algebras are stressed there, and the corresponding Kac-Kazhdan determinant formulas are compared in the appendix. 
Section 4 analyzes the structure of simple pole singularities in non-integrable
characters, when $\sigma$, the angular variable keeping track of the isospin of the states within an irreducible representation, tends to zero. It is found that, upon multiplying by a given modular form, the residues at the poles
are minimal $N=2$ superconformal characters, in complete parallel with 
a similar study in \cite{MPANDA} for $\hat{sl}(2,\C)$ characters, where the Virasoro
minimal characters emerge as residues at the poles.
In section 5, the branchings of $\hslc$ admissible characters at level $k=-\hf$
into $\hat{sl}(2,\C)$ characters at the same level are calculated, and the 
characters are shown to provide 
a finite representation of the modular group. 
\section{The characters and supercharacters of $\hslck$.}
\renewcommand{\theequation}{2.\arabic{equation}}
\setcounter{equation}{0}

The Lie superalgebra $A(1,0)\equiv sl(2|1;\mathbb C)$ has rank two and its simple roots
can be chosen to be fermionic. The root
diagram (Fig.1) can be represented in a 2-dimensional Minkowski space with the
 fermionic roots $\pm \alpha_1$ and $\pm \alpha_2$ in the lightlike directions (see \cite{BT} for notations),
\vskip .5cm

\begin{picture}(151,180)(017,435)
\thicklines
\multiput(180,603)(16.64190,-22.18921){8}{\line( 3,-4){  8.667}}
\put(305,436){\vector( 3,-4){0}}
\put(180,603){\vector(-3, 4){0}}
\multiput(305,603)(-16.64190,-22.18921){8}{\line(-3,-4){  8.667}}
\put(180,436){\vector(-3,-4){0}}
\put(305,603){\vector( 3, 4){0}}
\put(117,520){\vector(-1, 0){  0}}
\put(117,520){\vector( 1, 0){251}}
\put(243,520){\circle*{6}}
\put(315,435){\makebox(0,0)[lb]{\smash{${\bf \alpha_2}$}}}
\put(315,600){\makebox(0,0)[lb]{\smash{${\bf \alpha_1}$}}}
\end{picture}
\vskip 1cm
\centerline{Figure 1:\it{ The root diagram of $A(1,0)$.}}
\vskip 1cm

\noindent

The currents of the untwisted affine superalgebra $A(1,0)^{(1)} \equiv \hslc$ 
have the following Laurent expansions,
\begin{xalignat}{2}
J({\bf e}_{\pm(\alpha_1+\alpha_2)})(z)&=\sum _n J_n^{\pm} z^{-n-1}&\notag\\
J({\bf h}_-)(z)&=2\sum _n J_n^3 z^{-n-1}& 
J({\bf h}_+)(z)&= 2\sum_n U_n z^{-n-1}\notag\\
J({\bf e_{\pm \alpha_1}})(z) &= \sum _n j_n^{'\pm}z^{-n-1}&J({\bf e_{\pm \alpha_2}})(z) &= 
\sum _n j_n^{\pm}z^{-n-1}. 
\end{xalignat}
In terms of these Laurent modes, 
the commutation relations for $A(1,0)^{(1)}$ at arbitrary level $k$ are,
\begin{xalignat}{2}
[J_m^+,J_n^-]&= 2J_{m+n}^3+km \delta_{m+n,0}&{[}U_m,U_n{]}&= -\frac{k}{2}m \delta_{m+n,0}\notag\\
{[} J_m^3,J_n^{\pm} {]} &= \pm J_{m+n}^{\pm}&[J_m^3,J_n^3]&=\frac{k}{2}m \delta_{m+n,0}\notag\\
{[}J_m^{\pm},j_n^{'\mp}{]}&=\pm j_{m+n}^{\pm}&{[}J_m^{\pm},j_n^{\mp}{]}&=\mp j_{m+n}^{'\pm}\notag\\
{[}2J_m^3,j_n^{'\pm}{]}&=\pm j_{m+n}^{'\pm}&[2J_m^3,j_n^{\pm}]&=\pm j_{m+n}^{\pm}\notag\\
{[}2U_m,j_n^{'\pm}{]}&=\pm j_{m+n}^{'\pm}&[2U_m,j_n^{\pm}]&=\mp j_{m+n}^{\pm}\notag\\
\{j_m^{'+},j_n^{'-} \}&= (U_{m+n}-J_{m+n}^3)-km \delta_{m+n,0}&\notag\\
\{j_m^+,j_n^- \}&= (U_{m+n}+J_{m+n}^3)+km \delta_{m+n,0}&\{ j_m^{'\pm},j_n^{\pm} \} &= J_{m+n}^{\pm}.
\label{super}
\end{xalignat}
Also, the Sugawara energy-momentum tensor is given by,
\bea
T(z)&=& \frac{1}{2(k+1)} \{ 2(J^3)^2(z) - 2U^2(z) + J^+J^-(z) + J^-J^+(z)\nn
&& +j^{'+}j^{'-}(z) - j^{'-}j^{'+}(z) 
-j^+j^-(z) + j^-j^+(z) \}.
\label{suga}
\eea
Until otherwise stated, we shall work in the Ramond sector, where the
suffix of all generators is an integer.

As first discussed in \cite{KW88}, the level $k$ of an affine Lie (super)algebra
must be of the form
\begin{gather}
k+h^{\vee}=\frac{p}{u}, \qquad p,u \in \N \qquad \gcd(p,u)=1,
\end{gather}
for admissible representations to exist. 
In the above expression, 
$h^{\vee}$ is the dual Coxeter number of the Lie (super)algebra 
and $h^{\vee}=1$ for $sl(2|1;\mathbb C)$.
Setting $u=1$, the level is an integer ($k\in\zplus\defd\N\cup\{0\}$ when $h^{\vee}=1$), which
is a necessary condition for integrable representations.

In this paper, we identify a family of irreducible highest weight state (hws) representations
which are non-integrable but nevertheless  
 whose
characters provide a finite representation of the modular
group. This indicates that one can build rational conformal field theories based
 on $\super$ at fractional level.

The superalgebra $A(1,0)$ possesses two sets of Weyl inequivalent simple roots, for instance, $\{\alpha_1,\alpha_2\}$ and $\{ \alpha_1+\alpha_2, -\alpha_2\}$. Clearly, the definition of hws depends crucially on the choice of positive roots. In this paper, we have chosen $\alpha_1, \alpha_2$ and
$\alpha_1+\alpha_2$ as positive roots, and $|\Lambda \rangle$ is a hws when
\ben
j_0^{+'} |\Lambda \rangle=j_0^{+} |\Lambda \rangle=J_1^-|\Lambda \rangle =0.
\label{hws}\end{equation}
If instead, the positive roots are chosen to be $\alpha_1+\alpha_2,-\alpha_2$ and $\alpha_1$, the hws $|\Lambda '\rangle$ is defined by,
\ben
J_0^+|\Lambda'\rangle=j_0^-|\Lambda'\rangle=j_0^{+'}|\Lambda '\rangle=0 ,
\end{equation}
but the qualitative analysis of characters is unchanged.

An $\hslck$ hws $|\Lambda\rangle$ is further characterized by its isospin $\hf h_-$ and its hypercharge 
$\hf h_+$, \emph{i.e.},
\ben
J_0^3 |\Lambda \rangle = \half h_- |\Lambda \rangle, 
\qquad U_0 |\Lambda \rangle = \half h_+ |\Lambda \rangle,
\end{equation}
while its conformal weight, calculated from the Sugawara tensor \eqref{suga},
is given by,
\ben
h=\frac{1}{4(k+1)}(h_-^2-h_+^2).
\end{equation}
For some specific values of $h_-$ and $h_+$, dictated by the Kac-Kazhdan
determinant formula, the Verma module built on such a hws contains singular vectors.
 The identification of their quantum numbers and of their 
embedding patterns within the Verma module is of crucial importance in constructing
 irreducible characters. A classification of 
embedding diagrams for $\hslck$ singular vectors appearing in Verma modules
built on hws $|\Lambda\rangle$ whose isospin quantum number $h_-$ obeys the
constraint
\begin{gather}
h_-+ \frac{p}{u}m-n=0,\notag\\
0 \leqslant m \leqslant u-1, \qquad  0 \leqslant n \leqslant p-1,
\label{cond1}\end{gather}
was provided in \cite{BT}.

Of particular interest here is the class where the isospin $\hf h_-$ obeys
the condition \eqref{cond1} with $n=0$, but also where the hypercharge 
$\hf h_+$ is constrained by
\ben
h_--h_+=-2\frac{p}{u}m',\qquad m' \in \zplus, \qquad m'-m \leqslant 0.
\end{equation}
These two conditions on the hws quantum numbers $\hf h_-$ and $\hf h_+$ can be reformulated as
\begin{gather}
h_{-}+h_{+}+2(k+1)(m-m')=0,\notag\\
h_{-}-h_{+}+2m'(k+1)=0,
\label{class4}\end{gather}
where $ m,m'\in{\zplus}, 0\leqslant m'\leqslant m\leqslant u-1
\text{ and }k+1=\frac{p}{u}.$
They correspond to class IV in  \cite{BT}.

However, in order to obtain a rational conformal field theory, one is led
to consider, together with this class IV, a new class of representations
for which the hws has isospin given by \eqref{cond1} with $n=p$, and 
hypercharge $\hf h_+$ given by,
\ben
h_--h_+=2 \frac{p}{u}(m'+1),\qquad m' \in\zplus, \qquad m+m' \leqslant u-2.
\end{equation}
These latter two conditions are equivalent to the following constraints on $h_-$
and $h_+$,
\begin{gather}
h_{-}+h_{+}-2(k+1)(u-m-m'-1)=0,\notag\\
h_{-}-h_{+}-2(k+1)(m'+1)=0,
\label{class5}
\end{gather}
where $m,m'\in\zplus, 0\leqslant m+m'\leqslant u-2 \text{ and }
k+1=\frac{p}{u}.$
This new class is labelled class V.

The embedding diagrams for classes IV and V have the same pattern, given in Fig. 2.
 However, the quantum numbers of singular vectors in the two classes are different.
We reproduce the class IV data and give the new, class V data in the following tables
(where $a\in\zplus.)$

\begin{figure}
\centering
\epsfig{file=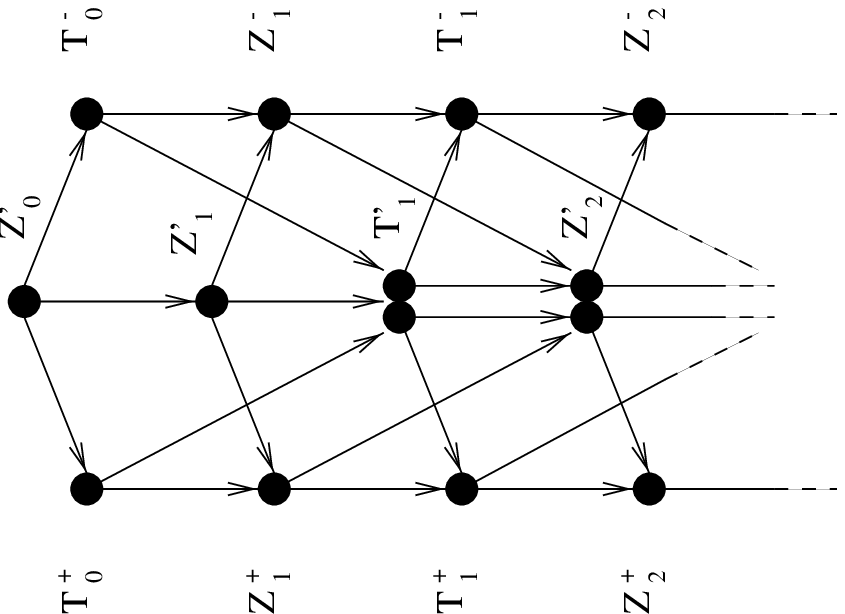,height=3.54in,width=4.5in,clip=,angle=270}
\caption{\it{The embedding diagram for classes IV and V.}}
\end{figure}

In the notation of \cite{BT}, 
$ Z'_{0}$ is the bosonic hws. $Z'_{a+1}$ and 
$T'_{a+1}$ are bosonic singular vectors while
$T^{+}_{a}, Z^{+}_{a+1},
T^{-}_{a}$ and $Z^{-}_{a+1}$ are fermionic singular vectors.

If the hws $Z'_{0}$ 
has conformal weight $h$, isospin $\half h_{-}$ and hypercharge $\half h_{+}$
satisfying the criteria in \eqref{class4}, then the 
quantum numbers of the singular vectors are given in Table 1 and Table 2.

\setlength{\extrarowheight}{6pt}
\[
\begin{tabular}{|>{$}c<{$}!{\vrule width 2pt}>{$}c<{$}|} \hline   
\text{Family}    & \text{Conformal weight}             \\ \hline\hline
Z'_{a}           & h+a^{2}pu-apm                   \\ 
T'_{a+1}         & h+(a+1)^{2}pu+(a+1)pm           \\ 
Z^{-}_{a+1}      & h+m'-m+(a+1)^{2}pu+(a+1)(u-pm)  \\ 
T^{-}_{a}        & h+m'+a^{2}pu+a(u+pm)            \\ 
Z^{+}_{a+1}      & h-m'+(a+1)^{2}pu+(a+1)(u-pm)    \\ 
T^{+}_{a}        & h+m-m'+a^{2}pu+a(u+pm)          \\ 
\hline
\end{tabular}
\]
\vskip 1cm
\centerline{{\bf {\rm Table~1}}:\it ~Conformal~ weights~ of ~class~ IV~ Ramond~singular ~vectors.  }
\vskip 1cm
\[
\begin{array}{|c!{\vrule width 2pt}c|c|} \hline
\text{Family}    & 2\times\text{Isospin}      & 2\times\text{Charge}     \\ \hline\hline 
Z'_{a}           & 2ap-m(k+1)                 & h_{+}                    \\
T'_{a+1}         & -2(a+1)p-m(k+1)            & h_{+}                    \\
Z^{-}_{a+1}      & 1+2(a+1)p-m(k+1)           & h_{+}-1                  \\
T^{-}_{a}        & -1-2ap-m(k+1)              & h_{+}-1                  \\
Z^{+}_{a+1}      & 1+2(a+1)p-m(k+1)           & h_{+}+1                  \\
T^{+}_{a}        & -1-2ap-m(k+1)              & h_{+}+1                  \\ 
\hline
\end{array}
\]
\vskip 1cm
\centerline{{\bf {\rm Table~2.}}\it Isospin~and~Charge~of~class~IV~Ramond ~singular~ vectors.}
\vskip 1cm
If instead, the hws $Z'_0$ has conformal weight $h$, isospin $\half h_{-}$ and hypercharge $\half h_{+}$
satisfying the criteria in \eqref{class5}, then the quantum numbers
 of the singular vectors are given in Table 3 and Table 4, with $M=u-2-m-m'$ and $M'=m'$.

\[
\begin{array}{|c!{\vrule width 2pt}c|} \hline
\text{Family}    & \text{Conformal weight}                \\ \hline\hline
Z'_{a}           & h+a^{2}pu-ap(M+M'+2)                   \\
T'_{a+1}         & h+(a+1)^{2}pu+(a+1)p(M+M'+2)                   \\
Z^{-}_{a+1}      & h-M'-1+(a+1)^{2}pu+(a+1)(u-p(M+M'+2))   \\
T^{-}_{a}        & h+M+1+a^{2}pu+a(u+p(M+M'+2))          \\
Z^{+}_{a+1}      & h-M-1+(a+1)^{2}pu+(a+1)(u-p(M+M'+2))  \\
T^{+}_{a}        & h+M'+1+a^{2}pu+a(u+p(M+M'+2))           \\
\hline
\end{array}
\]
\vskip 1cm
\centerline{{\bf {\rm Table~3}}:\it ~Conformal~ weights~of~class~V~Ramond~singular~ vectors.}
\vskip 1cm
\[
\begin{array}{|c!{\vrule width 2pt}c|c|} \hline
\text{Family}    & 2\times\text{Isospin}      & 2\times\text{Charge}     \\ \hline\hline
Z'_{a}           & -2ap+(M+M'+2)(k+1)         & h_{+}                    \\
T'_{a+1}         & 2(a+1)p+(M+M'+2)(k+1)      & h_{+}                    \\
Z^{-}_{a+1}      & -2(a+1)p-1+(M+M'+2)(k+1)   & h_{+}-1                  \\
T^{-}_{a}        & 2ap+1+(M+M'+2)(k+1)        & h_{+}-1                  \\
Z^{+}_{a+1}      & -2(a+1)p-1+(M+M'+2)(k+1)   & h_{+}+1                  \\
T^{+}_{a}        & 2ap+1+(M+M'+2)(k+1)        & h_{+}+1                  \\
\hline
\end{array}
\]
\vskip 1cm
\centerline{{\bf {\rm Table~4}}:\it ~Isopin~and~Charge~of~class~V~Ramond~singular~vectors. }
\vskip 1cm
The information above is sufficient to derive the Ramond sector $\hslck$ characters
for the representations in classes IV and V , when the
level is given by $k=-1+\frac{p}{u}, p, u \in \N\text{ and } \gcd(p,u)=1$.

In the Ramond sector, the Kac-Weyl denominator
for $\super$ corresponds to the free action of all negatively moded generators of the algebra,
 together with $J_0^-, j_0^-$ and $j_0^{'-}$. It reads,
\begin{multline}
(F^R(\sigma, \nu, \tau))^{-1} =\\[2mm]
\prod^{\infty}_{n=1}\frac{(1-q^{n})^{2}(1-zq^{n})(1-z^{-1}q^{n-1})}
{(1+z^{\frac{1}{2}}\zeta^{\frac{1}{2}}q^{n})(1+z^{-\frac{1}{2}}\zeta^{\frac{1}{2}}q^{n-1})
(1+z^{\frac{1}{2}}\zeta^{-\frac{1}{2}}q^{n})(1+z^{-\frac{1}{2}}\zeta^{-\frac{1}{2}}q^{n-1})},
\label{factorR}
\end{multline}
where,
\begin{align}
 & q\defd\exp(2\pi \mathrm{i}\tau), \quad \tau \in \mathbb C \quad   
      \im(\tau)>0\Rightarrow|q|<1,\notag\\
 & z\defd\exp(2\pi \mathrm{i}\sigma), \quad \sigma\in\mathbb C,\notag\\
 & \zeta\defd\exp(2\pi \mathrm{i}\nu), \quad \nu\in\mathbb C.
\end{align}
The complex variables
$q,z,\zeta$ respectively keep track of the conformal weight, isospin and charge  of the states
in the Verma module. Using the standard technique of subtracting off the submodules generated from the singular vectors,
whose quantum numbers are given in Tables 1 - 4, one obtains the 
Ramond characters for irreducible, admissible representations of class IV as,
\begin{gather}
\chi^{R,IV,\super}_{h_-^R,h_+^R}(\sigma,\nu,\tau)=q^{h^{R}}z^{\frac{1}{2} h_{-}^{R}}
\zeta^{\frac{1}{2}h_{+}^{R}} F^R(\sigma, \nu, \tau)\times \notag\\[3mm]
\sum_{a\in\Z}q^{a^{2}pu+apm}z^{-ap}\frac{1-q^{2ua+m}z^{-1}}{
(1+q^{au+m'}z^{-\frac{1}{2}}\zeta^{-\frac{1}{2}})(1+q^{au+m-m'}z^{-\frac{1}{2}}\zeta^{\frac{1}{2}})},\label{rad4}
\end{gather}
where $ m,m'\in\zplus , 0\leqslant m'\leqslant m\leqslant u-1,$ and
\begin{gather}
h_-^R=-m(k+1),\notag\\
h_+^R=(2m'-m)(k+1).
\label{qnr4}
\end{gather}

On the other hand,
the Ramond characters for irreducible, admissible representations of class V are,
\begin{gather}
\chi^{R,V,\super}_{h_-^R,h_+^R}(\sigma,\nu,\tau)=q^{h^{R}}z^{\frac{1}{2} h_{-}^{R}}
\zeta^{\frac{1}{2}h_{+}^{R}} F^R(\sigma, \nu, \tau) \times \notag\\[3mm]
\sum_{a\in\Z}q^{a^{2}pu+ap(M+M'+2)}z^{ap}\frac{1-q^{2au+M+M'+2}z}{(1+q^{au+M+1}z^{\frac{1}{2}}\zeta^{-\frac{1}{2}})
(1+q^{au+M'+1}z^{\frac{1}{2}}\zeta^{\frac{1}{2}})},\label{rad5}
\end{gather}
where $M,M'\in\zplus , 0\leqslant M+M'\leqslant u-2,$ and 
\begin{gather}
h_-^R=(M+M'+2)(k+1),\notag\\
h_+^R=(M-M')(k+1).
\label{qnR5}
\end{gather}

The Neveu--Schwarz sector of the theory may be obtained from the Ramond sector using
 a variety of spectral flows, corresponding to several isomorphisms of the
algebra $A(1,0)^{(1)}$. For instance, the following `tilde' 
generators, 
\begin{gather}
\tilde{J}_{n\pm 2\beta}^{\pm}=J_n^{\pm},\notag\\
\tilde{j}_{n\pm \beta \mp\alpha}^{\pm}=j_n^{\pm},\notag\\
\tilde{j}_{n\pm \beta \pm \alpha }^{\pm '}=j_n^{'\pm},\notag\\
\tilde{J}_n^3=J_n^3+k\beta \delta _{n,0},\notag\\
\tilde{U}_n=U_n-k\alpha \delta_{n,0},\notag\\
\tilde{L}_n=L_n+2\beta J_n^3 -2\alpha U_n -k(\alpha^2-\beta^2)\delta_{n,0},
\end{gather}
obey the same algebra as the original Ramond generators for any real values
of the parameters $\alpha$ and $\beta$. In particular, if the pair
$(\alpha , \beta ) = (0, \pm \hf) \text{ or } (\pm \hf ,0)$, one relates the Ramond sector to the 
Neveu-Schwarz sector of the theory. The first two  
values of $(\alpha , \beta)$ correspond to isospin spectral flows, while the
latter two correspond to hypercharge spectral flows. Although these
flows may be used to derive the Neveu-Schwarz characters of $\hslc$ from the
Ramond characters given above, it should be emphasized that the isospin 
flows do not always map highest weight state representations to highest 
(or lowest) weight representations. This is due to the fact that some admissible Ramond
 representations at fractional level have an infinite
number of states at grade zero. 
There exists, however, an inner automorphism of the algebra $sl(2|1;\mathbb C)$
which changes the sign of all nonzero roots in the root diagram and which corresponds to
 the following isomorphism of the affine version of the algebra,
\begin{xalignat}{2}
J_{n}^{3,NS}&=-J_{n}^{3,R}+\half k\delta_{n,0}, & L_{n}^{NS}&=L_{n}^{R}-J_{n}^{3,R}+\quarter k\delta_{n,0},\notag\\
U_{n}^{NS}&=U_{n}^{R}, & J^{\mp, NS}_{n\pm1}&=J^{\pm, R}_{n},\notag\\
j^{'\mp,NS}_{n\pm\textstyle{1\over2}}&=j^{\pm,R}_{n}, & j^{\mp,NS}_{n\pm\textstyle{1\over2}}&=j^{'\pm,R}_{n}.
\label{flow}
\end{xalignat}
The central generator $k$ remains unchanged.
The above isomorphism maps a Ramond hws to a \ns hws, whose quantum numbers
are given in terms of the quantum numbers of the Ramond hws by,
\begin{align}
h^{NS}&=h^{R}-\half h^{R}_{-}+\quarter k,\notag\\
\half h_{-}^{NS}&=-\half h^{R}_{-}+\half k,\notag\\
\half h_{+}^{NS}&=\half h_{+}^{R}.\label{changedqnos}
\end{align}
The \ns characters may then be obtained from the Ramond characters by using
the spectral flow \eqref{flow} as follows, 
\begin{align}
\chi^{NS,\super}_{h_-^{NS},h_+^{NS}}
(\sigma,\nu,\tau)&\defd\text{tr}\exp\bigl(2\pi \mathrm{i}(\tau L_{0}^{NS}+\sigma
J_{0}^{3,NS}+\nu U_{0}^{NS})\bigr)\notag\\
&=\text{tr}\exp\bigl(2\pi \mathrm{i}(\tau(L_{0}^{R}-J_{0}^{3,R}+\quarter k)+\sigma(-J_{0}^{3,R}+\half k)
+\nu U_{0}^{R}\bigr)\notag\\
&=q^{\quarter k}z^{\half k}\text{tr}\exp\bigl(2\pi\mathrm{i}(\tau L^{R}_{0}-(\sigma+\tau)J^{3,R}_{0}
+\nu U^{R}_{0}\bigr)\notag\\
&=q^{\quarter k}z^{\half k}\chi^{R,\super}_{h_-^R,h_+^R}(-\sigma-\tau,\nu,\tau).
\end{align}

Defining the infinite product,
\begin{multline}
F^{NS}(\sigma, \nu, \tau) = \\[3mm]
\prod^{\infty}_{n=1}\frac{(1+z^{\frac{1}{2}}\zeta^{\frac{1}{2}}q^{n-\hf})
(1+z^{-\frac{1}{2}}\zeta^{\frac{1}{2}}q^{n-\hf})
(1+z^{\frac{1}{2}}\zeta^{-\frac{1}{2}}q^{n-\hf})
(1+z^{-\frac{1}{2}}\zeta^{-\frac{1}{2}}q^{n-\hf})}
{(1-q^{n})^{2}(1-zq^{n})(1-z^{-1}q^{n-1})},
\label{factorNS}
\end{multline}
the Neveu--Schwarz characters for class IV irreducible, admissible 
representations of $\super$ therefore read,
\begin{gather}
\chi^{NS,IV,\super}_{h_-^{NS},h_+^{NS}}(\sigma,\nu,\tau)=q^{h^{NS}}z^{\frac{1}{2} h_{-}^{NS}}
\zeta^{\frac{1}{2} h_{+}^{NS}} F^{NS}(\sigma, \nu, \tau) \times \notag\\[3mm]
\sum_{a\in\Z}q^{a^{2}pu+ap(1+m)}z^{ap}\frac{1-q^{2au+1+m}z}{
(1+q^{au+m'+\frac{1}{2}}z^{\frac{1}{2}}\zeta^{-\frac{1}{2}})
(1+q^{au-m'+m+\frac{1}{2}}z^{\frac{1}{2}}\zeta^{\frac{1}{2}})},\label{nsad4}
\end{gather}
where $m,m'\in\zplus \text{ and }0\leqslant m'\leqslant m\leqslant u-1$. 
Also, the \ns characters for class V irreducible, admissible representations are,
\begin{gather}
\chi^{NS,V,\super}_{h_-^{NS},h_+^{NS}}(\sigma,\nu,\tau)=q^{h^{NS}}z^{\frac{1}{2} h_{-}^{NS}}
\zeta^{\frac{1}{2} h_{+}^{NS}} F^{NS}(\sigma, \nu, \tau) \times \notag\\[3mm]
\sum_{a\in\Z}q^{a^{2}pu+ap(M+M'+1)}z^{-ap}\frac{1-q^{2au+M+M'+1}z^{-1}}
{(1+q^{au+M+\frac{1}{2}}z^{-\frac{1}{2}}\zeta^{-\frac{1}{2}})
(1+q^{au+M'+\frac{1}{2}}z^{-\frac{1}{2}}\zeta^{\frac{1}{2}})},
\notag\\
\label{nsad5}
\end{gather}
where $M,M'\in\zplus \text{ and } 0\leqslant M+M'\leqslant u-2.$
%
%
%
%
%

It is now straightforward to derive the supercharacters by shifting 
$\sigma\longrightarrow\sigma+1$ in the expressions for the characters
\eqref{rad4},\eqref{rad5},\eqref{nsad4},\eqref{nsad5} and dividing by
$e^{2\pi i\frac{1}{2}h_{-}^{R\,{\rm or}\,NS}}$.
This amounts to the same as inserting the operator $(-1)^F$ in the formal definition of
a character because the isospin of a fermionic singular vector
is offset from that of a bosonic singular vector by $\hf$. 
 In the Ramond sector, one obtains the class IV 
supercharacters as,
\begin{gather}
S\chi^{R,IV,\super}_{h_-^R,h_+^R}(\sigma,\nu,\tau)=q^{h^{R}}z^{\frac{1}{2} h_{-}^{R}}\zeta^{\frac{1}{2} 
h_{+}^{R}}F^{R}(\sigma+1,\nu,\tau)\times\notag\\[3mm]
\sum_{a\in\Z}q^{a^{2}pu+apm}z^{-ap}\frac{1-q^{2au+m}z^{-1}}
{(1-q^{au+m'}z^{-\frac{1}{2}}\zeta^{-\frac{1}{2}})(1-q^{au+m-m'}z^{-\frac{1}{2}}\zeta^{\frac{1}{2}})},\label{srad4}
\end{gather}
where $ m,m'\in\zplus \text{ and }0\leqslant m'\leqslant m\leqslant u-1,$
 while the class V Ramond supercharacters are,
\begin{gather}
S\chi^{R,V,\super}_{h_-^R,h_+^R}(\sigma,\nu,\tau)=q^{h^{R}}z^{\frac{1}{2} h_{-}^{R}}\zeta^{\frac{1}{2}  
h_{+}^{R}}F^{R}(\sigma+1,\nu,\tau)\times\notag\\[3mm]
\sum_{a\in\Z}q^{a^{2}pu+ap(M+M'+1)}z^{ap}\frac{1-q^{2au+M+M'+2}z}
{(1-q^{au+M+1}z^{\frac{1}{2}}\zeta^{-\frac{1}{2}})(1-q^{au+M'+1}z^{\frac{1}{2}}\zeta^{\frac{1}{2}})},\label{srad5}
\end{gather}
where $ M,M'\in\zplus \text{ and } 0\leqslant M+M'\leqslant u-2.$ Upon putting
$\sigma=\nu=0$ into the \emph{nonsingular} (see section 4) supercharacters
we discover that all such supercharacters reduce to unity (class IV) or zero (class V).
These numbers may be interpreted as Witten indices. In the case of class V, although the Witten
index vanishes, the supersymmetry is unbroken.

In the \ns sector, the class IV supercharacters are,
\begin{gather}
S\chi^{NS,IV,\super}_{h_-^{NS},h_+^{NS}}(\sigma,\nu,\tau)=q^{h^{NS}}z^{\frac{1}{2}h_{-}^{NS}}
\zeta^{\frac{1}{2}h_{+}^{NS}}F^{NS}(\sigma+1,\nu,\tau)\times\notag\\[3mm]
\sum_{a\in\Z}q^{a^{2}pu+ap(m+1)}z^{ap}\frac{1-q^{2au+m+1}z}{(1-q^{au+m'+\frac{1}{2}}z^{\frac{1}{2}}\zeta^{-\frac{1}{2}})
(1-q^{au+m-m'+\frac{1}{2}}z^{\frac{1}{2}}\zeta^{\frac{1}{2}})},
\end{gather}
where $m,m'\in\zplus \text{ and } 0\leqslant m'\leqslant m\leqslant u-1$, while the class V Neveu-Schwarz supercharacters are,
\begin{gather}
S\chi^{NS,V,\super}_{h_-^{NS},h_+^{NS}}(\sigma,\nu,\tau)=q^{h^{NS}}z^{\frac{1}{2}h_{-}^{NS}}
\zeta^{\frac{1}{2}h_{+}^{NS}}F^{NS}(\sigma+1,\nu,\tau)\times\notag\\[3mm]
\negthickspace\negthickspace\negthickspace\negthickspace
\sum_{a\in\Z}q^{a^{2}pu+ap(M+M'+1)}z^{-ap}\frac{1-q^{2au+M+M'+1}z^{-1}}
{(1-q^{au+M+\frac{1}{2}}z^{-\frac{1}{2}}\zeta^{-\frac{1}{2}})
(1-q^{au+M'+\frac{1}{2}}z^{-\frac{1}{2}}\zeta^{\frac{1}{2}})},
\end{gather}
where $M,M'\in\zplus \text{ and } 0\leqslant M+M' \leqslant u-2.$

We have thus obtained the Ramond sector $\hslc$ characters at 
fractional level $k$, when the hws quantum numbers are given by \eqref{class4}
and \eqref{class5}. The Neveu-Schwarz characters were obtained by spectral flow.
Should one insist on an integer level $k$, the only possible hws quantum numbers are $h_+=h_-=0$, stemming from \eqref{class4} with $u=1$, and the corresponding representation in the Ramond sector is the vacuum representation. Remarkably, this integrable vacuum representation
and other integrable, non-vacuum representations
arising from class I (in the classification of \cite{BT}) have
characters identical to those of the $N=4$ superconformal algebra at the same
level $k$. We shall discuss this
relationship now.
\section{Integrable representations and their characters.}
\renewcommand{\theequation}{3.\arabic{equation}}
\setcounter{equation}{0}
Let $\hat{L}(\Lambda)$ be a highest weight module over $A(1,0)^{(1)}$.
If $\alpha_1\text{ and }\alpha_2$ are the simple roots of $A(1,0)$, we
can parametrize $\Lambda$ by,
\ben
\Lambda = (\hf (h_-+h_+) \alpha _1 + \hf (h_--h_+) \alpha _2, k, 0).
\end{equation} 
This module is integrable if and only if one of the following conditions is satisfied : 
$h_- \in \N$ and $h_+$ is unconstrained,
or $h_-=h_+=0.$ We also require
the level $k \in \zplus$ with $k\geqslant h_-$\cite{KW94}. Since $k$ is to be an integer, one
should specialize to the value
$u=1$ in \eqref{cond1}. A straightforward analysis of the hws quantum numbers
in classes I, II, III, IV and V reveals that integrable representations
 occur in all classes but 
III and V. It is a remarkable fact that the integrable characters of class I
 are the same as the massive $N=4$ 
characters discussed in \cite{ET881}, while the unique integrable representation
in class IV (which is the vacuum representation)
is the same as the massless $N=4$ representation at isospin $\ell=0$ (see \cite{ET881}).
 Before we establish these relations with the characters of the $N=4$ superconformal algebra,
let us stress that
  the $\super$ integrable characters just mentioned obey the ``non-degeneracy'' conditions,
\begin{gather}
\hf (h_--h_+) \neq 0 \Rightarrow \hf (h_--h_+) 
\text{ is not divisible by } k+1,\notag\\
\hf (h_-+h_+) \neq 0 \Rightarrow \hf (h_-+h_+) \text{ is not divisible by } k+1,
\end{gather}
and that the class IV integrable vacuum character agrees  with an explicit formula obtained by
Kac and Wakimoto in \cite{KW94}.

We will now indicate how to relate integrable characters of classes I and 
IV to massive and massless $N=4$ characters respectively. The integrable
 characters in class I correspond to
 representations whose highest weight $\Lambda$
has arbitrary charge $\hf h^R_+$, isospin,
\ben
\hf h^R_-=\hf n,\qquad 1 \leqslant n\leqslant k=p-1,\qquad n \in \N,
\end{equation}
and conformal weight
\begin{equation}
 h^R_{\super}=\frac{1}{4(k+1)}((h^R_-)^2-(h^R_+)^2).
\end{equation}
 Using the embedding diagram and quantum numbers of class I singular vectors given in \cite{BT},
one easily derives the following $\super$ Ramond sector, integrable characters,
\begin{gather}
\chi^{R,I,\super}_{h_-^R,h_+^R}(\sigma,\nu,\tau)=q^{h^{R}}z^{\frac{1}{2} h_{-}^{R}}
\zeta^{\frac{1}{2}h_{+}^{R}} F^R(\sigma, \nu, \tau) \times \notag\\[3mm]
\sum_{a\in\Z}q^{a^{2}(k+1)+ah^R_-}(z^{a(k+1)}-z^{-a(k+1)-h^R_-})\label{rad1}.
\end{gather}
Direct comparison with the massive $N=4$ characters given in \cite{ET881} shows that, 
once the latter are multiplied by the ``Casimir'' factor $q^{-\frac{c}{24}}=q^{-\frac{k}{4}}$,
 once the conformal weights 
of the hws are related by $h^R_{N=4}=h^R_{\super}+\frac{k}{4},$
and once one identifies 
\ben
h_-^{R}=2\ell,\quad 2\pi\sigma = \theta \quad\text{and}\quad 2\pi\nu = \varphi ,
\label{identif}
\end{equation}
the two expressions are identical up to a factor $\zeta ^{\hf h_+}$. The
reason for this is that
in the $N=4$ characters, the variable $y$ keeps track of the $U(1)$ charge of
the supersymmetry generators, and it was assumed that the hws had charge zero. Setting $h_+=0$,
one then may write the Ramond sector, integrable characters of class I as,
\begin{multline}
\chi^{R,I,\super}_{h_-^R,0}(\sigma,\nu,\tau)=\\[2mm]
\frac{1}{\eta (\tau)} 
\biggl(\chi_0^{\iso _1}(\sigma, \tau) \chi_{\hf}^{\iso _1}(\nu, \tau) +
\chi_{\hf}^{\iso _1}(\sigma, \tau) \chi_0^{\iso _1}(\nu, \tau)\biggr)
\chi_{\hf (h^R_--1)}^{\iso _{k-1}}(\sigma ,\tau).
\end{multline}
In the above, the integrable $\iso _k$ characters are given by,
\ben
\chi_{\ell}^{\iso _k}(\sigma,\tau)=\frac{\vartheta_{2\ell+1,k+2}(\sigma,\tau)
-\vartheta_{-2\ell-1,k+2}(\sigma,\tau)}{\vartheta_{1,2}(\sigma,\tau)-\vartheta_{-1,2}(\sigma,\tau)},~~~~~0 \le \ell \le \frac{k}{2}, ~2\ell \in \Z,
\label{su2}
\end{equation}
where the generalised theta functions (see e.g. \cite{KBOOK})
\ben
\vartheta _{m,k}(\sigma,\tau,w)\defd e^{2\pi ikw}\sum_{a\in\Z}e^{2\pi i\tau k(a+\frac{m}{2k})^2}
e^{2\pi i\sigma k(a+\frac{m}{2k})}= e^{2\pi ikw}~\vartheta_{m,k}(\sigma,\tau),
\label{theta}
\end{equation}
are evaluated at $w=0$, and where $\eta(\tau)$ is 
Dedekind's function,
\ben
\eta(\tau)\defd q^{\textstyle{1\over24}}\prod^{\infty}_{n=1}(1-q^{n})=\vartheta_{1,6}(0,\tau)
-\vartheta_{5,6}(0,\tau).
\label{dedekind}
\end{equation}
The Neveu-Schwarz characters are obtained by spectral flow and read,
\begin{multline}
\chi^{NS,I,\super}_{h_-^{NS},0}(\sigma,\nu,\tau)=\\[2mm]
\frac{1}{\eta (\tau)} 
\biggl(\chi_{\hf }^{\iso _1}(\sigma, \tau) \chi_{\hf}^{\iso _1}(\nu, \tau) +
\chi_0^{\iso _1}(\sigma, \tau) \chi_0^{\iso _1}(\nu, \tau)\biggr)
\chi_{\hf h_-^{NS}}^{\iso _{k-1}}(\sigma ,\tau).
\label{massive}
\end{multline}
These characters have the following S-transform,
\begin{multline}
\chi^{NS,I,\super}_{h_-^{NS},0}(\frac{\sigma}{\tau},\frac{\nu}{\tau},\frac{-1}{\tau})=\\[2mm]
(-i\tau)^{-\hf} e^{-\frac{i\pi k\sigma^2}{\tau}}
e^{-\frac{i\pi \nu ^2}{\tau}}\sum_{h_-^{'NS}=0}^{k-1} S_{h_-^{NS},h_-^{'NS}} \chi^{NS,I,\super}_{h_-^{'NS},0}(\sigma,\nu, \tau),
\end{multline}
where,
\ben
 S_{h_-^{NS},h_-^{'NS}} =
\sqrt{\frac{2}{k+1}} \sin\biggl(\pi\frac{(h_-^{NS}+1)(h_-^{'NS}+1)}{k+1}\biggr).
\end{equation}

We now show that the massless, Ramond sector character of the vacuum representation of the $N=4$
 superconformal algebra 
studied in \cite{ET881}
is identical to the class IV Ramond sector vacuum character of $\hslck$ when $k$ is an integer.
 The latter is the
 expression \eqref{rad4} when $u=1$ and $m=m'=0$,\emph{i.e.},
\begin{multline}
\chi^{R,IV,\super}_{0,0}(\sigma,\nu,\tau)=\\[2mm]
F^R(\sigma,\nu,\tau) \times
\sum_{a\in\Z}q^{a^{2}p}z^{-ap}\frac{1-q^{2a}z^{-1}}{
(1+q^{a}z^{-\frac{1}{2}}\zeta^{-\frac{1}{2}})(1+q^{a}z^{-\frac{1}{2}}\zeta^{\frac{1}{2}})}.\label{rint}
\end{multline}
From \cite{ET881}, we have the following expression for the massless $N=4$
unitary characters at integer level $k$ and isospin $\ell =0, \hf,...,\frac{k}{2}$,
\begin{multline}
{\rm ch}^{R}_{0}(k,\ell;\sigma, \nu, \tau)=
q^{\frac{1}{4}k}F^{R}(\sigma ,\nu, \tau)\times\\[2mm]
\sum_{m\in\Z}\biggl(
\frac{z^{(k+1)m+\ell}q^{(k+1)m^{2}+2\ell m}}{(1+\zeta ^{\hf}z^{-\hf}q^{-m})(1+\zeta^{-\hf}z^{-\hf} 
q^{-m})}
-\frac{z^{-(k+1)m-\ell}q^{(k+1)m^{2}+2\ell m}}{(1+\zeta^{\hf}z^{\hf}q^{-m})(1+\zeta^{-\hf}z^{\hf} 
q^{-m})}\biggr),\\[2mm]
 \label{n=4char}
\end{multline}
where we used again the identifications \eqref{identif}.

Multiplying the numerator and denominator of the second term of the sum in \eqref{n=4char}
by $q^{2m}z^{-1}$, while relabeling $m\rightarrow-m$ in the first term and setting  $\ell=0$, we get,
\ben
\sum_{a\in\Z}q^{(k+1)a^{2}}z^{-(k+1)a}\frac{1-q^{2a}z^{-1}}{(1+q^{a}z^{-\frac{1}{2}}\zeta^{\frac{1}{2}})
(1+q^{a}z^{-\frac{1}{2}}\zeta^{-\frac{1}{2}})}.
\end{equation}
With $k+1=p$ this is identical to  the sum in \eqref{rint}. Thus the class IV vacuum
character at integer level $k$ is identical to the massless $N=4$ vacuum character at the same level
{\em i.e.},
\begin{equation}
\chi^{R,IV,\super}_{0,0}(\sigma,\nu,\tau)=q^{-\frac{1}{4}k}{\rm 
ch}_{0}^{R}(k,0;2\pi\sigma,2\pi\nu),
\end{equation}
where $q^{-\frac{1}{4}k}$ is the Casimir factor.
Similarly, putting $u=1,\,m=m'=0$ into \eqref{nsad4} yields the following expression for the
\ns sector integrable character,
\begin{multline}
\chi^{NS,IV,\super}_{k,0}(\sigma,\nu,\tau)=q^{\frac{k}{4}}z^{\frac{k}{2}}
F^{NS}(\sigma,\nu,\tau) \\[2mm]
\times\sum_{a\in\Z}q^{a^{2}p+ap}z^{ap}\frac{1-q^{2a+1}z}{
(1+q^{a+\frac{1}{2}}z^{\frac{1}{2}}\zeta^{-\frac{1}{2}})(1+q^{a+\frac{1}{2}}z^{\frac{1}{2}}\zeta^{\frac{1}{2}})},\label{nsint}
\end{multline}
and one easily shows that,
\ben
\chi ^{NS,IV,\super}_{k,0}(\sigma ,\nu,\tau)=q^{-\frac{1}{4}k}{\rm ch}_{0}^{NS}(k,\half k;2\pi\sigma ,2\pi\nu),
\label{massless}
\end{equation}
in total agreement with the spectral flow.

We  now recall the behaviour of the character in
\eqref{massless} under the modular transformation 
$S: \tau \rightarrow -\frac{1}{\tau}$. In \cite{ET882}, the decomposition of this character
at level $k=1$ into $\iso_1$ characters was given as,
\begin{multline}
\chi^{NS,IV,\super}_{k=1,0}(\sigma,\nu =0,\tau) =\\[2mm]
\biggl(\frac{-\chi_{\hf}^{\iso _1}(\tau)}
      {(\chi_0^{\iso _1}(\tau))^2+(\chi_{\hf}^{\iso _1}(\tau))^2} 
     + h_3(\tau) \chi_0^{\iso _1}(\tau)\biggr)\chi_0^{\iso _1}(\sigma,\tau)\\[2mm]
+\biggl(\frac{\chi_0^{\iso _1}(\tau)}
       {(\chi_0^{\iso _1}(\tau))^2+(\chi_{\hf}^{\iso _1}(\tau))^2} 
     + h_3(\tau) \chi_{\hf}^{\iso _1}(\tau)\biggr)\chi_{\hf}^{\iso _1}(\sigma,\tau),
\end{multline}
where,
\ben
h_3(\tau)=\frac{1}{\eta(\tau) \theta_3(\tau)} 
\sum_a \frac{q^{a^2/2-1/8}}{1+q^{m-1/2}},
\end{equation}
and 
\ben
h_3(\tau) +h_3(-1/\tau) = \frac{1}{\eta(\tau)} \int _{-\infty}^{\infty} 
d\alpha \frac{q^{\frac{\alpha^2}{2}}}{2\cosh \pi \alpha}
\end{equation}
is the Mordell integral \cite{MORD}.
It is a straighforward exercise to show that
\begin{gather}
\chi^{NS,IV,\super}_{k=1,0}(\sigma =0,\nu =0,\frac{-1}{\tau})=
-\chi^{NS,IV,\super}_{k=1,0}(\sigma=0,\nu =0,\tau) \notag\\ +
\int _{-\infty}^{\infty}
d\alpha \frac{q^{\frac{\alpha^2}{2}}}{2\cosh \pi \alpha} \times 
\chi^{NS,I,\super}_{k=1,0}(\sigma=0,\nu =0,\tau) 
\end{gather}
where
\ben
\chi^{NS,I,\super}_{k=1,0}(\sigma=0,\nu =0,\tau) =
\frac{1}{\eta(\tau)}{((\chi_0^{\iso _1}(\tau))^2+(\chi_{\hf}^{\iso _1}(\tau))^2)}, 
\end{equation}
according to \eqref{massive}.

We end this section by noting, interestingly, that the striking identity
between integrable $\hslck$ and $N=4$ characters is deeply rooted in the 
similar nature of singular vectors in the two theories. A close inspection
of the generalised Malikov-Feigin-Fuchs (MFF) construction for singular vectors
given in \cite{BT} for $\hslck$ in the Ramond sector and partially in \cite{ET881} for $N=4$
reveals that there is a one-to-one correspondence between  the set of generators
$\{j_0^-,j_0^{-'},J_0^-,J_{-1}^+\}$ in $\hslck$ 
and the set $\{G_0^1,\bar{G}_0^2,T_0^-,T_{-1}^+\}$ in $N=4$. In particular,
it is obvious that the singular vectors in massive $N=4$ Verma modules
based on a hws $|\Lambda'\rangle$ defined by
\ben
G_0^2|\Lambda'\rangle=\bar{G}_0^1|\Lambda'\rangle=T_{1}^-|\Lambda'\rangle=T_0^+|\Lambda'\rangle=0,
\end{equation}
and the singular vectors in class I $\hslck$ Verma modules based on a hws defined by \eqref{hws}, have the same analytic expressions
once the above correspondence is implemented. One can also establish a
similar dictionary between the singular vectors of the Ramond massless vacuum 
module and those of the unique class IV integrable $\hslck$ module. For instance, the state $G_0^1|\Lambda'\rangle$ (resp. $\bar{G}_0^2|\Lambda'\rangle$)
becomes singular in the $N=4$ massless case and corresponds to the state
$j_0^-|\Lambda \rangle$ (resp. $j_0^{-'}|\Lambda\rangle$) in class IV $\hslck$.
Although the currents in the two theories have non-matching conformal spin,
they have the same isospin and $U(1)$-charge quantum numbers, which implies that
the singular vectors have the same quantum numbers, a fact that can indeed be checked
by direct inspection of the Kac-Kazhdan determinant formulas of the two algebras
(see appendix).

Whether the relation between $N=4$ and $\hslck$ goes beyond coincidence of
character formulas is an open question. One should analyze the correlation
functions in the two theories, a task beyond the scope of this paper, but certainly
facilitated by the knowledge of the MFF representation of singular vectors.
\section{Character singularities.}
\renewcommand{\theequation}{4.\arabic{equation}}
\setcounter{equation}{0}

We now return to the class IV and class V $\hslck$ characters presented in section 2,
 and analyse their
 behaviour in the limit $\sigma \rightarrow 0$. It turns out, as we will see below,
 that some characters
 develop a simple pole in this limit, while others are 
regular at $\sigma =0$.
We find that upon multiplying by a 
certain modular form, the residues at the pole are (non)unitary, minimal $N=2$ \scon 
characters \cite{DOB}.
We can use these residues to identify the branching functions of $\hslck$ characters into $\iso _k$
 characters, which
makes the discussion of modular properties of characters in the next section much easier.
In \cite{MPANDA}, Mukhi and Panda had already remarked that some $\iso$ characters for
 admissible representations
(which were worked out in \cite{KW88}) developed a simple pole in a certain limit.
 Upon multiplication by $\eta^{2}$, the residues at these poles
were found to be the Virasoro characters in the unitary and non--unitary
minimal series. The results in \cite{MPANDA} and our analysis are especially interesting
 in the light of the 
relationship between $\iso _k$ and the Virasoro algebra and between $\super$ and the $N=2$
 \scon algebra
through quantum hamiltonian reduction.

Simple poles arise in the factor $F^R(\tau,\sigma,\nu)$ (see \eqref{factorR}) or
 $F^{NS}(\tau,\sigma,\nu)$ (see \eqref{factorNS}) of the $\hslck$ characters at $\sigma=0$.
 The following lemmas identify which characters are
singular at $\sigma =0$.
We shall see that this simple pole is cancelled by the vanishing of the sums in the characters
 only for 
special values of $m$ in the case of class IV and special values of $M+M'+1$ in the case of class V.
%
%
%
\begin{Lem}
The \r characters of class IV (resp. class V)
are nonsingular at $\sigma=0$ iff $m=0$ (resp. $M+M'+1=u-1$).
\end{Lem}
{\bfseries Proof}.
We prove the lemma only for class IV but the proof for class V is identical.
The sum from \eqref{rad4} is,
\[
\sum_{a\in\Z}q^{a^{2}pu+apm}z^{-ap}\frac{1-q^{2ua+m}z^{-1}}{
(1+q^{u+m'}z^{-\frac{1}{2}}\zeta^{-\frac{1}{2}})(1+q^{au+m-m'}z^{-\frac{1}{2}}\zeta^{\frac{1}{2}})}
\]
where $z\defd\exp(2\pi i\sigma)$.
Then when $\sigma$ is small the sum becomes,
\[
\sum_{a\in\Z}q^{a^{2}pu+apm}\frac{1-q^{2au+m}-2\pi i\sigma\bigl(ap+q^{2ua+m}(1+ap)\bigr)+O(\sigma^{2})}
{f(\sigma;a)}
\]
where the denominator $f(\sigma;a)$ is defined as,
\begin{multline*}
f(\sigma;a)\defd(1+q^{au+m-m'}\zeta^{\frac{1}{2}})(1+q^{au+m'}\zeta^{-\frac{1}{2}})\\
-i\pi\sigma(q^{au+m-m'}\zeta^{\frac{1}{2}}
+q^{au+m'}\zeta^{-\frac{1}{2}}+2q^{2au+m})+O(\sigma^{2}).
\end{multline*}
We can pair up each term in the sum with the one that has the equal but opposite value of $a$.
 Then the sum becomes,
\[
\half\biggl\{\sum_{a\in\Z}q^{a^{2}pu+apm}\frac{1-q^{2ua+m}-2\pi i\sigma\bigl(ap+q^{2ua+m}(1+ap)\bigr)
+O(\sigma^{2})}
{f(\sigma;a)}+(a\rightarrow-a)\biggr\}\notag
\]
\begin{multline*}
=\half\biggl\{\sum_{a\in\Z}q^{a^{2}pu}\frac{1-q^{2ua}-2\pi i\sigma\bigl(ap+q^{2ua}(1+ap)\bigr)
+O(\sigma^{2})}
{f(\sigma;a)}+\\
\sum_{a\in\Z}q^{a^{2}pu}\frac{1-q^{-2ua}-2\pi i\sigma\bigl(-ap+q^{-2ua}(1-ap)\bigr)+O(\sigma^{2})}
{f(\sigma;-a)}\biggr\}
\end{multline*}
iff $m=0$. Then multiplying numerator and denominator of the second term by 
$q^{2au}$, we see that the last line vanishes at $\sigma=0$.\hfill$\square$           
\begin{Cor}
Integrable characters are nonsingular at $\sigma=0$.
\end{Cor}
{\bfseries Proof}.
We set $u=1$ to obtain integrable characters in class IV.            
Since $0\leqslant m\leqslant u-1$, $u=1\Rightarrow m=0$.\hfill$\square$        

Similar arguments show that class IV \ns characters are nonsingular only when $m=u-1$.
However, the type of argument above does not prove that the class V \ns characters are ever 
nonsingular.
In the example of the next section we shall see that the class V \ns character that appears is 
singular.
We conjecture that the class V \ns characters are {\em all\/} singular.

The finite dimensional algebra $sl(2|1)$ is a subalgebra of the \r
sector $\super$. The next lemma,
when combined with Lemma 1, shows 
that the class IV \r sector characters of the affine algebra
are nonsingular iff the highest weight module $L(\Lambda_h)$ of the finite dimensional algebra is
finite dimensional.
\begin{Lem}
Let $\Lambda_{h}=(\half h_{-},\half h_{+})$ be the Ramond highest weight of $sl(2|1)$
where,
\[
\half h_{-}=-\frac{mp}{2u}\quad \text{and}\quad \half h_{+}=\frac{p}{2u}(2m'-m).
\]
Then the highest weight module $L(\Lambda_h)$ is finite dimensional iff $m=0$.
\end{Lem}
{\bfseries Proof}.
Let $\Pi=\{\alpha_{1}+\alpha_{2},-\alpha_{1}\}$ be a basis of simple roots for $sl(2|1)$.
$\alpha_{1}+\alpha_{2}$ is an even root. $\alpha_{1}$ and $\alpha_{2}$ are isotropic
odd roots. $\alpha_{1}\centerdot\alpha_{2}=1$ so, $(\alpha_{1}+\alpha_{2})^2=2.$
 The scalar product on the roots is computed with a Minkowski signature
metric. Between them, Kac \cite{K77} and Cornwell \cite{CORN} prove
\footnote{Kac proves sufficiency, Cornwell, necessity.}
that the graded representation  defined by $\Lambda_{h}$ is finite dimensional iff
the numerical mark $n_{1}$ (corresponding to $\alpha_{1}+\alpha_{2}$) is a 
non--negative integer, {\em i.e.,} iff,
\[
\Lambda_{h}\centerdot(\alpha_{1}+\alpha_{2})\in\zplus.
\]
Now, if we take $\alpha_{1}=\hf(1,1)\text{ and }\alpha_{2}=\hf(1,-1)$ then
\[
\Lambda_{h}=-\frac{mp}{u}\alpha_{1}+\frac{m'p}{u}(\alpha_{1}-\alpha_{2})
\]
and,
\[
\Lambda_{h}\centerdot(\alpha_{1}+\alpha_{2})=-\frac{mp}{u}\in\zplus\Leftrightarrow m=0.
\square
\]

Although $\Lambda_{h}$ is different for class V, it is easy to see that the lemma holds there too
for $M+M'+1=u-1$.

We end this section by giving 
the residues of class IV and class V \ns and \r $\hslck$ characters at the pole $\sigma=0$.

One has,
\begin{gather}
\lim\begin{Sb}
        \sigma\rightarrow 0
     \end{Sb}
2\pi i\sigma\,\chi^{NS,\super}_{h_-^{NS},h_+^{NS}}(\sigma,\nu,\tau) = 
\frac{\vartheta_{0,2}(\half\nu,\tau)+\vartheta_{2,2} (\half\nu,\tau)}{\eta^{3}(\tau)}
\chi^{NS,N=2}_{r,s}(\zeta^{\hf},q)\label{nsres}
\end{gather}
where the $N=2$ superconformal characters are given by,
\begin{gather}
\chi^{NS,N=2}_{r,s}(\zeta^{\hf},q)=\zeta^{\frac{(r-s)p}{2u}}q^{\frac{(rs-\frac{1}{4})p}{u}-\textstyle{3\over24}(1-\textstyle{2p\over u})}
 \prod_{n=1}^{\infty}\frac{(1+\zeta^{\hf}q^{n-\frac{1}{2}})(1+\zeta^{-\hf}q^{n-\frac{1}{2}})}
{(1-q^{n})^{2}}\times\notag\\[3mm]
\sum_{a\in\Z}q^{a^{2}up+ap(r+s)}\frac{1-q^{2ua+r+s}}{
(1+q^{ua+s}\zeta^{\hf})(1+q^{ua+r}\zeta^{-\hf})},\label{nschar}
\end{gather}
with central charge,
\begin{gather}
c=3(1-\frac{2p}{u}),\quad u\in\N\setminus\{1\},\quad p\in\N,\quad\gcd(u,p)=1.\label{c}
\end{gather}

In the above expression, we have changed notation as follows. For class IV, where $h_-^{NS}=m(k+1)+k$ and $h_+^{NS}= (2m'-m)(k+1)$, one has, $m=r+s-1,m'=r-\half$. For class V, where $h_-^{NS}=-(M+M'+2)(k+1)+k$
and $h_+^{NS}=(M-M')(k+1)$, one has, $M=r-\hf, M'=s-\hf$. From the ranges of $m,m',M\text{ and }M'$, we can
deduce ranges for $r\text{ and }s$ for each class and then comparing with Boucher, Friedan and Kent \cite{BFK}
we see that these $N=2$ characters-as-residues are unitary when $p=1$ and nonunitary when $p>1$.

In the Ramond sector, the residues of class IV and class V $\hslck$ characters
at the pole $\sigma=0$ are,
\begin{gather}
\lim\begin{Sb}
    \sigma\rightarrow0
    \end{Sb}
2\pi i\sigma\,\chi^{R,\super}_{h_-^R,h_+^R}(\sigma,\nu,\tau)= 
\frac{\vartheta_{1,2}(\half\nu,\tau)+\vartheta_{-1,2} (\half\nu,\tau)}{\eta^{3}(\tau)}
\chi^{R+,N=2}_{r,s}(\zeta^{\hf},q)\label{rres}\\[3mm]
\intertext{where,}
\chi^{R+,N=2}_{r,s}(\zeta^{\hf},q)=\zeta^{\frac{(r-s)p}{2u}}q^{\frac{rsp}{u}}(\zeta^{\frac{1}{4}}+\zeta^{-\frac{1}{4}})
\prod_{n=1}^{\infty}\frac{(1+\zeta^{\hf} q^{n})(1+\zeta^{-\hf}q^{n})}
{(1-q^{n})^{2}}\times\notag\\[3mm]
\sum_{a\in\Z}q^{a^{2}up+ap(r+s)}\frac{1-q^{2ua+r+s}}{
(1+q^{ua+s}\zeta^{\hf})(1+q^{ua+r}\zeta^{-\hf})}.\label{rchar}
\end{gather}

For class IV, where $h_-^R$ and $h_+^R$ are given by \eqref{qnr4}, 
$m=r+s\text{ and }m'=r$, but for class V, with $h_-^R$ and $h_+^R$ given by \eqref{qnR5}, $r=M+1\text{ and }s=M'+1.$
The $N=2$ characters derived from class V are unitary or not according to whether $p=1\text{ or }p>1$.
The $N=2$ characters derived from class IV are unitary when $m'>0\text{ and }p=1$, and are nonunitary 
otherwise.
\section{$\hslck$ admissible characters at level $k=-\hf$.}
\renewcommand{\theequation}{5.\arabic{equation}}
\setcounter{equation}{0}

In this section we show how some low level admissible $\super$ characters may be decomposed in a basis of
admissible $\iso _k$ characters, as $\iso _k$ is a subalgebra of $\super$. 
 From \cite{KW88}
or from \cite{MPANDA}, we have the following expression for the $\iso _k$ admissible characters,
\begin{equation}
\chi^{\iso _k}_{n,n'}(\sigma,\tau)
        =\frac{\vartheta_{b_{+},a}(\textstyle{\sigma\over u},\tau)
           -\vartheta_{b_{-},a}(\textstyle{\sigma\over u},\tau)}                                                                               {\vartheta_{1,2}(\sigma,\tau)-\vartheta_{-1,2}( \sigma,\tau)}, 
    \label{su2ad}
\end{equation}
where the level is parametrized as,
\begin{equation}
k=\frac{t}{u},~~~~~(t,u)=1,~~~u \in \N,~~t \in \Z,
\label{level}
\end{equation}
where
$ 0\leqslant n\leqslant 2u+t-2 \text{ and } 0\leqslant n'\leqslant u-1$ and,
\begin{equation}
b_{\pm}\defd u(\pm(n+1)-n'(k+2))\qquad a\defd u^{2}(k+2).
\end{equation}
Note that integer level requires $u=1$, which implies $n'=0$, and \eqref{su2ad} is then the well-known expression
\eqref{su2}, which is regular when $\sigma \rightarrow 0$. For $u \neq 1$
however, it is shown in \cite{MPANDA} that, at fixed level $k$, some 
$\iso$ characters develop a pole at $\sigma =0$, and that the residues
at the pole are the minimal Virasoro characters characterized by the
pair $(t+2u,u)$, multiplied by $\frac{1}{\eta ^2}$. On the other hand, we showed in the previous section that the residue at the pole $\sigma =0$ in Ramond $\hslck$ characters is the product of the modular function
$\frac{1}{\eta^2}$ by an $N=2$ minimal character and
a linear combination of $A_2$ characters, where $A_2$ is the rational torus algebra with one extra spin 2 generator \cite{DIJK}. One can therefore
speculate that the branching functions of $\hslck$ characters into $\iso _k$
characters involve, for $k+1=\frac{p}{u}$, the $A_2$ characters 
$\frac{\vartheta_{m,2}(\hf \nu, \tau)}{\eta(\tau)},~m=0,1,2,3$, multiplied 
by the ratio of $N=2$ and Virasoro minimal characters. From a `coset'
construction point of view, the central charge $c_{sl(2|1)}$ associated
with $\hslck$ through the Sugawara energy-momentum tensor is zero for any
level $k$, while the corresponding central charge for $\iso _k$ is
$c_{sl(2)}=\frac{3k}{k+2}$. One therefore expects the coset to have
central charge $c_{\rm coset}=c_{sl(2|1)}-c_{sl(2)}=-\frac{3k}{k+2}$. 
For $k+1=\frac{p}{u}$, one can rewrite,
\ben
c_{\rm coset}=1~+~3(1-\frac{2p}{u})~-~(1-\frac{6p^2}{u(p+u)}),
\end{equation}
where one can read off, on the right hand side, the central charges
of the torus algebra $A_2$, of minimal $N=2$ and of minimal Virasoro for the pair $(p+u,u)$.

When $p=1,~u=2$, $c_{\rm coset}=1$ is the $A_2$ central charge. The following detailed analysis of $\hslc$ at level $k=-\hf$ certainly confirms the above
heuristic argument. When $k=-\hf$, i.e. when
$t=-1\text{ and }u=2,$ 
there are four admissible $\iso _{-\hf}$ characters {\em viz,} $\chi_{0,0},\chi_{1,0},
\chi_{0,1},\chi_{1,1}$. The first pair are regular at $\sigma=0$ and the latter pair are singular there
\cite{MPANDA}.
On the other hand,
 there are three $\hslc _{-\hf}$ characters 
in class IV, with
$(h_-^R,h_{+}^{R})=
(0,0),(-\hf,\pm\hf)$, while there is only one in class V , with $(h_-^R,h_+^R)=(1,0)$.
As was mentioned above in section 4, not all the $\super$ characters develop a pole at $\sigma=0$.
For the case at hand, the class IV \r sector characters developing a pole have $(h_{-}^{R},h_{+}^{R}) =(-\hf,\pm\hf)$.
By contrast the only  singular class IV Neveu--Schwarz character has 
$(h^{NS}_{-},h_{+}^{NS})=(-\hf,0)$. In class V, the single \r character is nonsingular and the \ns character
 is singular.
Combined use of residues at the poles and spectral flow arguments allows for
an analytic derivation of the branching functions when decomposing
$\hslc$ characters at level $k=-\hf$ into $\iso$ characters
at the same level. 
 The $N=2$ and Virasoro characters--as--residues are evaluated at $c=0$ and so they appear as factors
 of unity on the {\sc RHS}. Letting $\mu$ be $0$ or $1$, we find in the \r sector, 
\begin{align}
\chi_{\mu,0}^{R,\hslcp}(\sigma,\nu,\tau)&=
\frac{1}{\eta(\tau)} \sum_{\rho =0}^1 \vartheta_{2\mu+2\rho,2}(\half\nu,\tau)\chi_{\rho,0}^{\isop}(\sigma,\tau),\notag\\
\chi_{-\hf,\mu-\hf}^{R,\hslcp}(\sigma,\nu,\tau)&=
\frac{1}{\eta(\tau)} \sum_{\rho =0}^1 \vartheta_{2\mu+2\rho+1,2}(\half\nu,\tau)\chi_{\rho,1}^{\isop}(\sigma,\tau), 
\label{rbranching}
\end{align}
 and for the \ns sector,
\begin{align}
\chi_{\mu - \frac{3}{2},0}^{NS,\hslcp}(\sigma,\nu,\tau)&=
\frac{1}{\eta(\tau)} \sum_{\rho =0}^1 \vartheta_{2\mu+2\rho,2}(\half\nu,\tau)\chi_{\rho,1}^{\isop}(\sigma,\tau),\notag\\
\chi_{0,\mu-\hf}^{NS,\hslcp}(\sigma,\nu,\tau)&=
\frac{1}{\eta(\tau)} \sum_{\rho =0}^1 \vartheta_{2\mu+2\rho-1,2}(\half\nu,\tau)\chi_{\rho,0}^{\isop}(\sigma,\tau)
\label{nsbranching}.
 \end{align}
The singular (resp. non-singular) $\hslcp$ characters are expanded into the singular (resp. non-singular)
$\isop$ characters. For the modular transformations of the Ramond and Neveu--Schwarz characters, it is useful to have
the Neveu--Schwarz supercharacters branched into $\iso$ characters too. This is easily obtained from the
branchings of the \ns characters in \eqref{nsbranching} above upon shifting $\sigma\rightarrow\sigma+1\text{
and dividing by } e^{2\pi i\frac{1}{2}h_{-}^{NS}}$. We obtain,
 \begin{align}
S\chi_{\mu -\frac{3}{2},0}^{NS,\hslcp}(\sigma,\nu,\tau)&=
\frac{1}{\eta(\tau)} \sum_{\rho =0}^1 (-1)^{\mu+\rho}
\vartheta_{2\mu+2\rho,2}(\half\nu,\tau)\chi_{\rho,1}^{\isop}(\sigma,\tau),\notag\\
S\chi_{0,\mu-\hf}^{NS,\hslcp}(\sigma,\nu,\tau)&=
\frac{1}{\eta(\tau)} \sum_{\rho =0}^1 (-1)^{\rho}
\vartheta_{2\mu+2\rho-1,2}(\half\nu,\tau)\chi_{\rho,0}^{\isop}(\sigma,\tau).
\label{snsbranching}
\end{align}

It is now straighforward to deduce the
behaviour of the \r and \ns $\hslc _{-\hf}$ characters under the modular group $PSL(2,\mathbb Z).$ First of all, let us recall the S modular transform of
the generalised theta functions \eqref{theta} (see \cite{KBOOK}),  
\ben
\vartheta_{m,k}(\frac{\nu}{\tau},\frac{-1}{\tau},v+\frac{\nu^{2}}{2\tau})=
{\sqrt{\frac{-i\tau}{2k}}}\thinspace
\sum_{r=0}^{2k-1}e^{\frac{-i\pi rm}{k}}\vartheta_{r,k}(\nu, \tau,v).
\end{equation}
From there, and using \eqref{dedekind}, one also has,
\ben
\eta(\frac{-1}{\tau})=\sqrt{-i\tau}\thinspace\eta(\tau).
\end{equation}
Furthermore, in view
of the definition \eqref{su2ad}, the S transform of the admissible $\iso_k$
characters are,
\ben
\chi ^{\iso_k}_{n,n'}(\frac{\sigma}{\tau},-\frac{1}{\tau})=
e^{-\frac{i\pi k \sigma^2}{\tau}}\sum_{\nu =0}^{2u+t-2} \sum_{\nu '=0}^{u-1}
S_{n\nu,n'\nu '} \chi^{\iso_k}_{\nu, \nu'}(\sigma,\tau),
\end{equation}
with
\begin{align}
&S_{n\nu,n'\nu'}=\sqrt{\frac{2}{u^2(k+2)}}(-1)^{n'(\nu +1)+(n+1)\nu'}
e^{-i\pi (k+2)n'\nu'} \sin\left[\frac{\pi (n+1)(\nu +1)}{k+2}\right].\notag\\
&
\end{align}

So, under the transformation $S:(\sigma,\nu,\tau)\rightarrow(\frac{\sigma}{\tau},\frac{\nu}{\tau}
,\frac{-1}{\tau})$, the Ramond and Neveu-Schwarz $\hslcp$ characters respectively transform as follows,
\begin{align}
&\chi^{R,\hslcp}_{\mu,0}
(\frac{\sigma}{\tau},\frac{\nu}{\tau},-\frac{1}{\tau})=-\hf 
e^{\frac{i\pi(\sigma^2-\nu^2)}{2\tau}}\notag\\
&\left[ \sum_{\rho =0}^{1} e^{2i\pi (\mu -\hf)\rho}
                  S\chi^{NS,\hslcp}_{\rho -\frac{3}{2},0}(\sigma,\nu,\tau)
- \sum_{\rho =0}^{1} e^{2i\pi \mu (\rho-\hf)}
                  S\chi^{NS,\hslcp}_{0,\rho -\hf}(\sigma,\nu,\tau)\right],\notag\\
&\notag\\
&\chi^{R,\hslcp}_{-\hf,\mu -\frac{1}{2}}
(\frac{\sigma}{\tau},\frac{\nu}{\tau},-\frac{1}{\tau})=\hf 
e^{\frac{i\pi  (\sigma^2-\nu^2)}{2\tau}}
\notag\\
&
\left[i \sum_{\rho =0}^{1} e^{2i\pi \mu \rho}
                  S\chi^{NS,\hslcp}_{\rho -\frac{3}{2},0}(\sigma,\nu,\tau)
- \sum_{\rho =0}^{1} e^{2i\pi (\mu -\hf)(\rho-\hf)}
                  S\chi^{NS,\hslcp}_{0,\rho -\hf}(\sigma,\nu,\tau)\right],
\notag\\
&
\end{align}
and

\begin{align}
&\chi^{NS,\hslcp}_{\mu -\frac{3}{2},0}
(\frac{\sigma}{\tau},\frac{\nu}{\tau},-\frac{1}{\tau})=\hf
e^{\frac{i\pi(\sigma^2-\nu^2)}{2\tau}}
\notag\\
&
\left[i \sum_{\rho =0}^{1} e^{2i\pi \mu \rho}
                  \chi^{NS,\hslcp}_{\rho -\frac{3}{2},0}(\sigma,\nu,\tau)
- \sum_{\rho =0}^{1} e^{2i\pi \mu (\rho-\hf)}
                  \chi^{NS,\hslcp}_{0,\rho -\hf}(\sigma,\nu,\tau)\right],\notag\\
&\notag\\
&\chi^{NS,\hslcp}_{0,\mu -\frac{1}{2}}
(\frac{\sigma}{\tau},\frac{\nu}{\tau},-\frac{1}{\tau})=-\hf
e^{\frac{i\pi ( \sigma^2-\nu^2)}{2\tau}}
\notag\\
&
\left[ \sum_{\rho =0}^{1} e^{2i\pi (\mu -\hf)\rho}
                  \chi^{NS,\hslcp}_{\rho -\frac{3}{2},0}(\sigma,\nu,\tau)
+ \sum_{\rho =0}^{1} e^{2i\pi (\mu -\hf)(\rho-\hf)}
                  \chi^{NS,\hslcp}_{0,\rho -\hf}(\sigma,\nu,\tau)\right].
\notag\\
&
\end{align}

As can be checked easily, when evaluated at $\sigma =\nu =0$, the matrices
$S^R$ and $S^{NS}$ of S transform are unitary and $
\bigl(S^{NS}\bigr)^4=1$.
Thus we see that the \r characters transform into the super \ns characters and the \ns characters transform into 
themselves under $S$, as expected.

Finally, let us  mention how these admissible characters transform under
$T:(\sigma,\nu,\tau)\rightarrow(\sigma,\nu,\tau+1)$.
 It is straightforward to see that,
\begin{align}
\chi^{R,\super}_{h_-^R,h_+^R}(\sigma,\nu,\tau+1)&=e^{2\pi ih^R}\chi^{R,\super}_{h_-^R,h_+^R}(\sigma,\nu,\tau)\notag\\
\chi^{NS,\super}_{h_-^{NS},h_+^{NS}}(\sigma,\nu,\tau+1)&=e^{2\pi ih^{NS}}S\chi^{NS,\super}_{h_-^{NS},h_+^{NS}}(\sigma,\nu,\tau).
\end{align}
for class IV and class V separately at any level $k$.

We have therefore verified that the affine superalgebra $\hslc$ at fractional level $k=-\hf$ allows
 for four irreducible admissible representations whose 
characters form a finite representation of the modular group.
We expect to obtain such admissible representations in classes IV and
V for any fractional level of the form $k+1=\frac{p}{u}, p,u \in \N, {\rm gcd}(p,u)=1$. 
However, the corresponding character formulas in the form given in section 2 are not
directly suited for the analysis of modular transformations, and work is in progress to rewrite them
 in terms of modular forms and functions.  
\section{Conclusion}
\renewcommand{\theequation}{6.\arabic{equation}}
\setcounter{equation}{0}
In this paper, we provide character formulas for integrable and admissible,
irreducible representations of the affine superalgebra $\hslck$. We explicitly show how the characters of the torus algebra $A_2$ arise as branching
functions of {\em admissible} $\hslc$ characters at level $k=-\hf$ into
$\iso $ at the same level. This rewriting of
admissible characters enables us to easily derive their modular transformations.
It is also argued that branching functions for other fractional values of the
level \eqref{levelp} should involve the product of $A_2$ characters with a ratio of minimal $N=2$ and Virasoro characters.

The most surprising result however is that one can identify the {\em integrable} $\hslck$ characters with those
of the superconformal $N=4$ algebra. Striking similarities between the zero mode
generators of the two algebras are responsible
for the existence of singular vectors with identical embedding structures and
identical quantum numbers in both theories, leading to identical character
formulas. These observations are reflected in the presence of common factors
in the Kac-Kazhdan determinant formulas, as explained in the appendix.
Whether or not the relation goes beyond matching character formulas is 
not clear, and we wish to conclude our work with further remarks and
speculations. For instance, it is possible, using the Malikov-Feigin-Fuchs construction for singular vectors, to derive differential equations obeyed by the correlators of free fields in both theories. These
equations should encode the conformal spins of the generators, which obviously
differ in $\hslc$ and $N=4$. However,
another intriguing coincidence is that the Coulomb gas representation of the $N=4$ SCA at level $k$ \cite{MATH,MATSUDA}, which can be obtained
by hamiltonian reduction from the affine $A(1/1)^{(1)}$ superalgebra 
\cite{ITO} but also \cite{TVPSS,HS1,HS2} from a representation of the doubly
extended $N=4$ SCA, uses three $SU(2)$ currents at level $k-1$, four free real fermions and a $U(1)$ current.
This is very similar to the current content of $\hslck$.  
The mismatch in conformal spins could possibly be resolved by the appropriate twisting of one algebra. The fact that $A(1,1)$ contains $A(1,0)$ as a subalgebra, that its affinisation provides the $N=4$ superconformal algebra
through hamiltonian reduction while $A(1,0)^{(1)}$ reduces to
the $N=2$ superconformal algebra may ultimately shed a new light on the 
relation between the ubiquitous $N=2$ string theory and its more complicated
$N=4$ counterpart.

\vskip 1cm
\noindent
{\bf Acknowledgments}
\vskip 2mm
M. Hayes acknowledges the British EPSRC for a studentship
and A. Taormina the British EPSRC for an advanced fellowship. Computer calculations were done with MapleV.

\vskip 1cm

\noindent
{\bf{Appendix}}
\vskip 2mm
\renewcommand{\theequation}{A.\arabic{equation}}
\setcounter{equation}{0}

We show how to identify common factors in the Kac-Kazhdan determinant formulas
of $N=4$ and $\hslc$ at level $k$. The former was conjectured in \cite{KR} and proven
in \cite{MATSUDA} using a Coulomb gas representation of $N=4$ and
screening charges. In the Ramond sector (setting $\rho =\eta=0$ in formula
(9) of \cite{KR}), it reads,
\begin{align}
&{\rm det} M_{\nu,s,c}(h,k,t) =\notag\\
 &A~\prod _{m,n >0} [f_{m,n}(h,k,t)]^{P^4(\nu-mn,s,c)}
\prod_{m \in \Z}\prod _{n >0}[g_{m,n}(h,k,t)]^{P^4(\nu-|m|n,s+{\rm sgn}(m)n,c)}
\notag\\
&\times\prod_{m \in \Z, \epsilon =\pm 1} [h_{m,\epsilon}(h,k,t)]^{P^4_{m,\epsilon}(\nu-|m|,s+\hf {\rm sgn}(m),c-\epsilon)},
\label{n=4}\end{align}
with $A$ a non-zero constant, 
\begin{align}
&f_{m,n}(h,k,t)=4t^2-(k+1)(4h-k)-((k+1)m+n)^2,\notag\\
&g_{m,n}(h,k,t)=2t~{\rm sgn}(m)+|m|(k+1)-n,\notag\\
&h_{m,\epsilon}(h,k,t)=4h-k+4m(2t+m(k+1)),
\end{align}
and ${\rm sgn}(0)=1$. The partition functions $P^4$ and $P^4_{m,\epsilon}$ are given
in \cite{KR}, but we shall only use the result that $P^4(0,0,0)=P^4_{m,\epsilon}
(0,0,0)=1$.
Also, in the above, $k$ is the level of the $N=4$ SCA in the
notations of \cite{ET881}, and the quantum numbers $h$ and $t$ are the conformal
weight and the isospin of the singular vector on which a reducible
Verma module is built. 
The Kac-Kazhdan determinant formula enables one to identify the quantum numbers
of singular vectors in this reducible Verma module, given some specific values of $h$ and $t$ obtained whenever (at least) one factor of the determinant
vanishes. 

Suppose now that $h$ and $t$ are the quantum numbers of an $N=4$
hws in a unitary representation, i.e.,
\ben
h \ge \frac{k}{4}~~\text{and}~~\frac{k}{2} \ge t \ge 0, ~~2t \in \Z ,
\end{equation}
and let $c$ be its $U(1)$ charge. The $g$-product in \eqref{n=4} corresponds,
in the $\hslck$ Kac-Kazhdan determinant formula \cite{K86,DOB2}, to the
factors,
\begin{align}
&\tilde{\phi}_n^{(0)}((\bar{\alpha}_1+\bar{\alpha}_2),0,m),\notag\\
&\tilde{\phi}_n^{(0)}(-(\bar{\alpha}_1+\bar{\alpha}_2),0,1+m),~~~~~~~~m \in \zplus,~n \in \N ,
\end{align}
in the notations of \cite{BT}, when the $\hslck$ hws has isospin $\hf h_-$
and hypercharge $\hf h_+$. So, whenever a factor of the $g$-product vanishes, i.e.
whenever the isospin quantum number $t$ of the hws  obeys,
\ben
2t~{\rm sgn}(m)+(k+1)|m|-n=0,~~~\text{for}~m \in \Z ~~\text{and}~n \in \N,
\end{equation}
there exists a singular vector in the reducible hws Verma module with conformal weight
$h+mn$, isospin $t-{\rm sgn}(m)n$ and charge $c$. Using the formula iteratively  provides the quantum numbers of singular vectors appearing in the Verma module built on a massive $N=4$ hws. These quantum numbers, as well as the corresponding embedding diagrams, are identical to those in the $\hslck$
theory, therefore showing how the massive $N=4$ characters at level $k$
coincide with the class I integrable $\hslck$ characters.

Now consider the
$h$-factors in \eqref{n=4} when the conformal weight of the
$N=4$ hws is $h=\frac{k}{4}$,
\ben
4m(2t+m(k+1)),~~~~~m \in \Z.
\end{equation}
The second factor corresponds to the factors,
\begin{align}
&\tilde{\phi}_n^{(1)}((\bar{\alpha}_i,0,m),\notag\\
&\tilde{\phi}_n^{(1)}(-(\bar{\alpha}_i,0,1+m),~~~~~~~~m \in \zplus,~i=1,2 ,
\end{align}
in \cite{BT}, when the $\hslck$ hws has hypercharge $\hf h_+=0$. So, whenever the $N=4$ hws quantum number 
$t$ obeys,
\ben
2t+(k+1)m=0,~~~~~m \in \Z,
\end{equation}
there exists a singular vector in the hws Verma module with conformal weight
$\frac{k}{4}+|m|$, isospin $t-\hf {\rm sgn}(m)$ and hypercharge $c \pm 1$. In the unitary domain however, the only possible isospin value for the hws is $t=0$. So, provided
that $c=0$, the vacuum massless $N=4$ character at level $k$ coincides with the vacuum integrable class IV $\hslck$ character.
\bibliographystyle{plain}

\begin{thebibliography}{10}

\bibitem{AHA}
G.~Aharony, O.~Ganor, N.~Sochen, J.~Sonnenschein, and S.~Yankielowicz.
\newblock { Physical states in $G/G$ models and 2d gravity. }.
\newblock {\em Nuclear Physics B}, 399:527, 1993.

\bibitem{BO}
M.~Bershadsky and H.~Ooguri.
\newblock { Hidden $Osp(N,2)$ symmetries in superconformal field theories.}
\newblock {\em Physics Letters B}, 229:374, 1989.

\bibitem{BFK}
W.~Boucher, D.~Friedan, and A.~Kent.
\newblock {Determinant Formulae and Unitarity...}
\newblock {\em Physics Letters B}, 172:316, 1986.

\bibitem{BOUW}
P.~Bouwknegt, J.~McCarthy, and K.~Pilch.
\newblock { Semi-infinite cohomology in conformal field theory and 2d gravity.
  }.
\newblock {\em Comm. Math. Phys.}, 145:541, 1992.

\bibitem{BT}
P.~Bowcock and A.~Taormina.
\newblock {Representation Theory of the Affine Lie Superalgebra
 {$sl(2|1;\mathbb C)$} at Fractional Level.}
\newblock hep-th 9605220. To appear in Comm. Math. Phys.

\bibitem{HS2}
C.G. Callan, J.A. Harvey, and A.~Strominger.
\newblock {Supersymmetric string solitons}.
\newblock hep-th 9112030.

\bibitem{HS1}
C.G. Callan, J.A. Harvey, and A.~Strominger.
\newblock {Worldbrane actions for string solitons}.
\newblock {\em Nuclear Physics B}, 367:60, 1991.

\bibitem{CORN}
J.~F. Cornwell.
\newblock {\em {Group Theory in Physics Vol. 3}}.
\newblock Academic Press, {First} edition, 1989.

\bibitem{DIJK}
R.~Dijkgraaf, C.~Vafa, E.~Verlinde, and H.~Verlinde.
\newblock {The operator algebra of orbifold models}.
\newblock {\em Communications in Mathematical Physics}, 123:485, 1989.

\bibitem{DOB}
V.~K. Dobrev.
\newblock {Characters of the Unitarizable Highest Weight Modules Over the $N=2$
  Superconformal Algebra}.
\newblock {\em Physics Letters B}, 186 number 1:43, 1987.

\bibitem{DOB2}
V.K. Dobrev.
\newblock {\em { Multiplets of Verma Modules over the $osp(2/2)^{(1)}$ super
  Kac-Moody algebra. in:Topological and Geometrical Methods in Field Theory,
  Proceedings, eds. J. Hietarinta and J. Westerholm (Espoo 1986)}}.
\newblock World Scientific, 1986.



\bibitem{DORZ}
M.~D{\"o}rrzapf.
\newblock {\em Superconformal Field Theories and their Representations}.
\newblock PhD thesis, University of Cambridge, 1995.

\bibitem{ET881}
T.~Eguchi and A.~Taormina.
\newblock {Character Formulas for the $N=4$ Superconformal Algebra}.
\newblock {\em Physics Letters B}, 200, number 3:315, 1988.

\bibitem{ET882}
T.~Eguchi and A.~Taormina.
\newblock {On the unitary representations of $N=2$ and $N=4$ superconformal
  algebras}.
\newblock {\em Physics Letters B}, 210:125, 1988.

\bibitem{EGAB}
W.~Eholzer and M.~Gaberdiel.
\newblock {Unitarity of Rational $N=2$ Superconformal Theories}.
\newblock DAMTP preprint 96-06.


\bibitem{FY}
J-B. Fan and M.~Yu.
\newblock {G/G gauged supergroup valued WZWN field theory. }.
\newblock hep-th 9304122.

\bibitem{HY}
H.L. Hu and M.~Yu.
\newblock {On the equivalence of noncritical strings and $G/G$ topological
  field theories.}
\newblock {\em Physics Letters B}, 289:302, 1992.

\bibitem{ITO}
K.~Ito, J-O. Madsen, and J-L. Petersen.
\newblock {Free field representations of extended superconformal algebras}.
\newblock {\em Nuclear Physics B}, 398:425, 1993.

\bibitem{K77}
V.~G. Kac.
\newblock {Lie Superalgebras}.
\newblock {\em Advances in Mathematics}, 26:8, 1977.

\bibitem{KBOOK}
V.~G. Kac.
\newblock {\em {Infinite dimensional Lie algebras}}.
\newblock Cambridge University Press, {Third} edition, 1990.

\bibitem{KW94}
V.~G. Kac and M.~Wakimoto.
\newblock {Integrable Highest Weight Modules over Affine Superalgebras and
  Number Theory}.
\newblock preprint hep--th 9407057.

\bibitem{KW88}
V.~G. Kac and M.~Wakimoto.
\newblock {Modular Invariant Representations of Infinite--Dimensional Lie
  Algebras and Superalgebras}.
\newblock {\em Proceedings of the National Academy of Sciences USA}, 85:4956,
  1988.

\bibitem{K86}
V.G. Kac.
\newblock {\em { Highest weight representations of conformal current algebras.
  in:Topological and Geometrical Methods in Field Theory, Proceedings, eds. J.
  Hietarinta and J. Westerholm (Espoo 1986)}}.
\newblock World Scientific, 1986.

\bibitem{KAR}
D.~Karabali and H.J. Schnitzer.
\newblock {BRST quantization of the gauged WZW action and coset conformal field
  theories. }.
\newblock {\em Nuclear Physics B}, 329:625, 1990.

\bibitem{KR}
A.~Kent and H.~Riggs.
\newblock {Determinant formulae for the $N=4$ superconformal algebras}.
\newblock {\em Physics Letters B}, 198:491, 1987.

\bibitem{MATH}
P.~Mathieu.
\newblock {Representation of the $SO(n)$ and $U(n)$ superconformal algebras via
  Miura transformations}.
\newblock {\em Physics Letters B}, 218:185, 1989.

\bibitem{MATSUDA}
S.~Matsuda.
\newblock {Coulomb gas representations and screening operators of the $N=4$
  superconformal algebras}.
\newblock {\em Physics Letters B}, 282:56, 1992.

\bibitem{MORD}
L.~J. Mordell.
\newblock {The definite integral $\int_{-\infty}^{\infty}
  \frac{\exp[ax^2+bx]}{\exp[cx+d]}dx$ and the analytic theory of numbers}.
\newblock {\em Acta Math.}, 61:323, 1933.

\bibitem{MPANDA}
S.~Mukhi and S.~Panda.
\newblock {Fractional--Level Current Algebras and the Classification of
  Characters}.
\newblock {\em Nuclear Physics}, B338:263, 1990.

\bibitem{SEMI}
A.~Semikhatov.
\newblock {The non-critical N=2 string is an $sl(2/1)$ theory }.
\newblock {\em Nuclear Physics B}, 478:209, 1996.

\bibitem{TVPSS}
A.~Sevrin, P.~Spindel, W.~Troost, and A.~van Proeyen.
\newblock {Complex structures on parallelized group manifolds and
  supersymmetric sigma models}.
\newblock {\em Physics Letters B}, 206:71, 1988.

\end{thebibliography}

\end{document}